\documentclass[final,10pt]{elsarticle}

\usepackage[top=2.5cm,bottom=2.5cm,left=2.5cm,right=2.5cm]{geometry}

\usepackage[switch]{lineno}
\usepackage{setspace}
\usepackage{hyperref}
\usepackage{multirow}
\usepackage{booktabs}
\usepackage{amsmath,amsfonts,amssymb,amscd,amsthm,xspace}
\modulolinenumbers[10]
\setpagewiselinenumbers

\journal{Nuclear Instrument and Methods in Physics Research A}

\begin{document}

\begin{frontmatter}

\title{Fabrication of quencher-free liquid scintillator-based, high-activity $^{222}$Rn calibration sources for the Borexino detector}

\author[myamericanaddress,myitalianaddress]{D. Bravo-Bergu\~no}

\author[Milano]{L. Miramonti}
\author[myamericanaddress]{P. Cavalcante}
\author[Napoli,Napoli2]{V. Roca}
\author[myamericanaddress]{S. Hardy}
\author[myamericanaddress]{R.B. Vogelaar}

\address[myamericanaddress]{Physics Department, Virginia Tech, 24061 Blacksburg, VA (USA)}
\address[myitalianaddress]{INFN Sezione Milano, Via Celoria 16, 20133 Milano (Italy)}
\address[Milano]{Universit\'a degli Studi di Milano, Via Celoria 16, 20133 Milano (Italy)}
\address[Napoli]{INFN Sezione Napoli, Naples (Italy)}
\address[Napoli2]{Dipartimento di Fisica, Universit\'a di Napoli Federico II, Naples (Italy)}

\begin{abstract}
A reliable and consistently reproducible technique to fabricate $^{222}$Rn-loaded radioactive sources ($\sim$0.5-1 kBq just after fabrication) based on liquid scintillator (LS), with negligible amounts of LS quencher contaminants, was implemented. This work demonstrates the process that will be used during the Borexino detectorÕs upcoming calibration campaign, where one or several $\sim$100 Bq such sources will be deployed at different positions in its fiducial volume, currently showing unprecedented levels of radiopurity. These sources need to fulfill stringent requirements of $^{222}$Rn activity, transparency to the radiations of interest and complete removability from the detector to ensure their impact on Borexino's radiopurity is negligible. Moreover, the need for a clean, undistorted spectral signal for the calibrations imposes a tight requirement to minimize scintillator quenching agents ("quenchers") to null or extremely low levels.
\end{abstract}

\end{frontmatter}

\linenumbers
\doublespacing

\section{Introduction}
\label{sec:intro}

The calibration campaign for the Borexino neutrino observatory that took place in 2007-10\cite{calib} employed, for the first time, a novel technique that enabled to obtain relatively high activity ($\mathcal{O}(100)$ Bq) $^{222}$Rn calibration sources suitable to be deployed inside the detector's extremely radiopure Inner Volume\cite{Hardy}. Building upon this development, the technique was optimized in preparation for the upcoming 2017 Borexino-SOX calibration campaign. In particular, this was done by precisely characterizing loading procedures, improving the fluid handling system's reliability and, most importantly, by demonstrating the reliable fabrication of sources with low levels of contaminants affecting the liquid scintillator's (LS) response to ionization through reductions of its intrinsic light yield (henceforth referred to as chemical quenchers or, for convenience, just "quenchers"; see Section~\ref{subsec:quenching})\cite{my_thesis}. This paper intends to provide a comprehensive overview of the source and technique designs, as well as the results from the fabricated specimens. 

This technique is suitable for a large number of low-background detector calibration purposes, despite being developed exclusively for Borexino. Indeed, $^{222}$Rn offers a unique mixture of calibration signals ($\alpha$, $\beta$ and $\gamma$, over broad areas inside the energy region of interest up to $\sim$3 MeV), owing to the cascading decays of its decay daughters --particularly, $^{218}$Po, $^{214}$Bi and $^{214}$Po. Pulse-shape discrimination (PSD) techniques, energy response, fiducialization and position reconstruction, fast coincidences ($^{214}$Bi-Po, for example) and ionization quenching effects\cite{long_paper} (involving the $dE/dx$ of different types of radiation, to be distinguished from chemical quenching induced by quenchers) can therefore be reliably and quickly studied with $^{222}$Rn-based sources. At the same time, LS-based $^{222}$Rn sources with negligible chemical quencher concentrations offer the same scintillation response than the surrounding medium, with little to no spectral distortion thanks to the simple and optimized containment vial, even for short-range radiation. Additionally, the source is completely removable and, following the proper procedure described in \cite{calib}, has proven not to leave long-term radioactive contaminants in Borexino's active volume.

\section{Motivation}
\label{sec:motivation}

\subsection{The Borexino Neutrino Observatory}
\label{subsec:borex}

The Borexino liquid scintillator detector is devoted to performing high-precision neutrino observations. In particular, it is optimized to study the low energy part of the solar neutrino spectrum in the sub-MeV region, having the precision measurement of the $^7$Be solar neutrinos as its design objective. Borexino has succeeded in performing high-precision measurements of all the major components of the solar neutrino spectrum (first direct detections of \textit{pp}\cite{pp}, \textit{pep}\cite{pep}, $^7$Be\cite{7Be}, and lowest (3 MeV) threshold observation of $^8$B\cite{8B}), as well as in reaching the best available limit in the subdominant CNO solar neutrino rate\cite{8B}, with just the DAQ time of 767 days comprising its first dataset \textit{Phase 1} from 2007-10, as well as a more recent high-precision determinations of the aforementioned major solar neutrino fluxes using new techniques and enlarged statistics from the post-LS purification phase (\textit{Phase 2})\cite{wideband} \cite{new8B}. Geoneutrinos have also been detected by Borexino with high significance (5.9$\sigma$\cite{geo}) thanks to the extremely clean $\overline{\nu}$ channel. Results on searches for new particles, (anti)neutrino sources and rare processes like \cite{sterile_old}, \cite{antinu_sources}, \cite{pauli_trans}, \cite{e_decay}, \cite{axions} are expected to gain even more relevance during the \textit{Short-distance neutrino Oscillations with boreXino} (SOX) phase of the experiment, where a $\overline{\nu}_e$ generator will be placed in close proximity to the detector, in order to probe for anomalous oscillatory behaviors and unambiguously check for the expected experimental signatures along the phase space light sterile neutrinos might lie in\cite{SOX} \cite{Giunti}.

These results were possible thanks to the unprecedented, extremely radio-pure conditions reached in the active section of the detector --achieved thanks to a combination of ultra-clean construction and fluid-handling techniques as well as dedicated scintillator purification campaigns\cite{purif}. Detailed detector response determination was made possible thanks to very successful internal calibration campaigns\cite{calib} which did not disturb the uniquely radio-pure environment. 

\begin{figure}[ht]
\centering\small\includegraphics[width=0.5\linewidth]{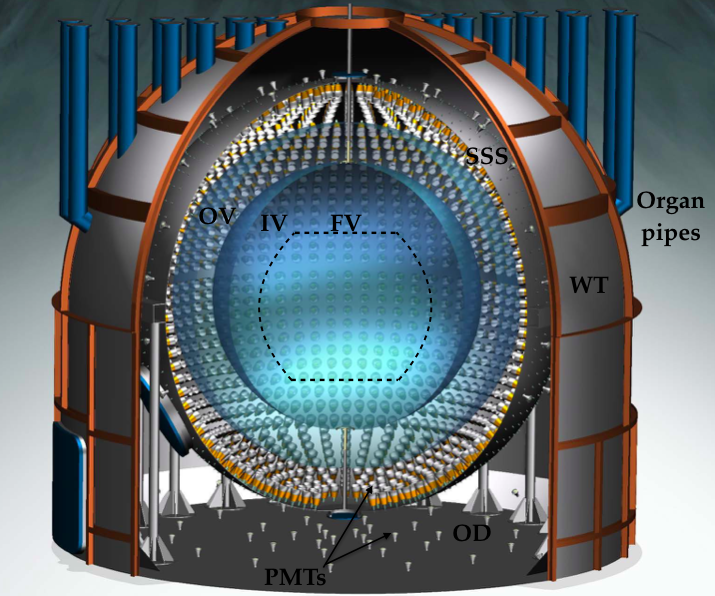}
\caption{The Borexino neutrino observatory, with its main structures annotated. See full text for details.}
\label{fig:BX}
\end{figure}

Borexino, located in the Hall C of the Gran Sasso National Laboratories' (LNGS) underground facilities (3,800 m w.e.), measures solar neutrinos via their interactions with a 278 tonnes target of organic LS. This ultrapure LS (pseudocumene --PC-- or 1,2,4-trimethylbenzene solvent with 1.5 g/l 2,5-diphenyloxazole --PPO-- scintillating solute) is contained inside a thin transparent spherical nylon Inner Vessel (IV) of 8.5 m diameter. Solar neutrinos are detected by measuring the energy and position of electrons scattered by neutrino-electron elastic interactions. The scintillator promptly converts the kinetic energy of electrons by emitting photons, which are detected and converted into electronic signals (photoelectrons or p.e.) by 2,212 photomultipliers (PMT) mounted on a concentric 13.7 m-diameter stainless steel sphere (SSS, see Figure~\ref{fig:BX}). A software-defined, analysis-dependent Fiducial Volume (FV) is established inside the IV. The volume between this inner nylon vessel and the SSS is filled with 889 tonnes of ultra pure, non scintillating fluid called "buffer" (PC+2-3 g/L dimethylphthalate or DMP) acting as a radiation shield for external gamma rays and neutrons. A second, larger nylon sphere (Outer Vessel --OV--, 11.5 m diameter) prevents radon and other radioactive contaminants from the PMTs and SSS from diffusing into the central sensitive volume of the detector, and segments the Inner and Outer Buffers (IB and OB). The SSS is immersed in a 2,100-tonne Water Tank (WT) acting as a \v{C}erenkov detector detecting residual cosmic $\mu^{\pm}$.

\subsection{$^{222}$Rn calibrations}
\label{subsec:222rn}

\begin{figure}[ht]
\centering\small\includegraphics[width=1\linewidth]{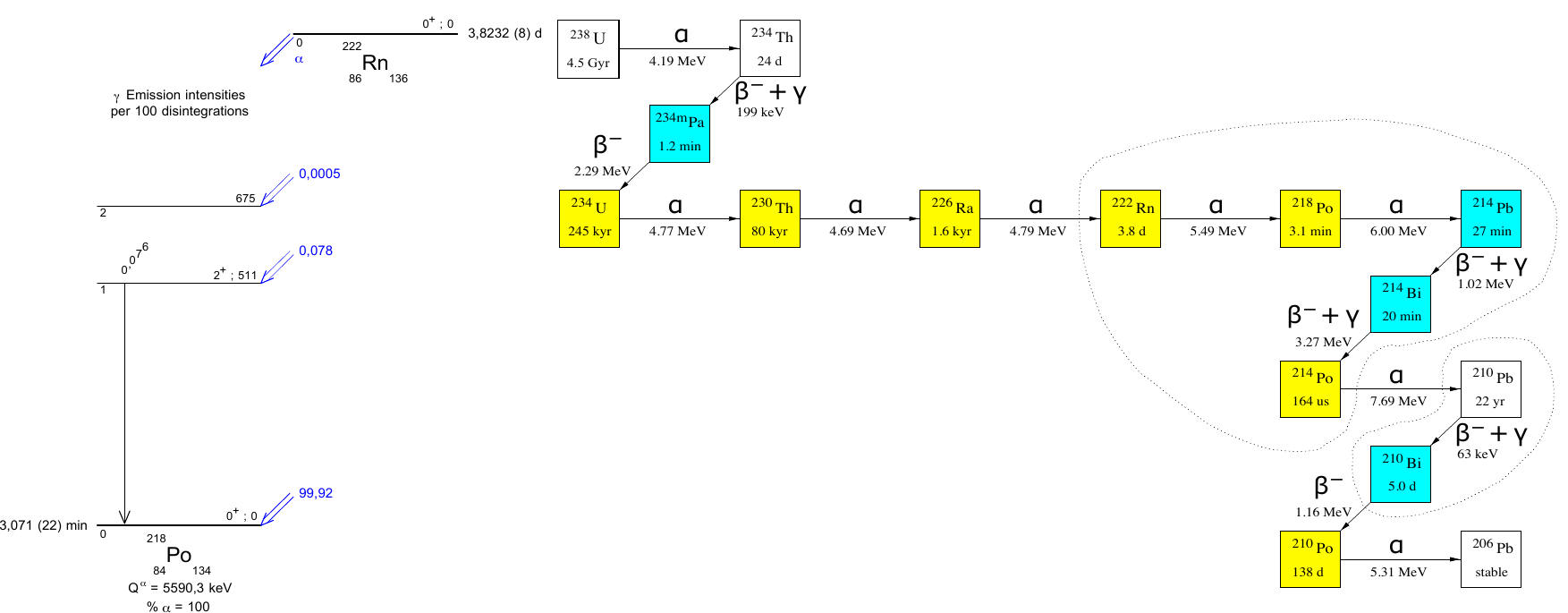}
\caption{$^{222}$Rn $\alpha$-decay (left side) and decay chain from its daughers (right side), inscribed in the general $^{238}$U chain to which they belong. Encircled in dotted lines are the components which are likely to hold secular equilibrium in Borexino. $^{210}$Po is also near equilibrium with $^{210}$Pb-Bi at the current Borexino age. Yellow(blue)-shaded isotopes are $\alpha$($\beta$)-emitters with a spectral endpoint above 250 keV at the ionization-quenched (-unquenched) visible energy in the detector, respectively. The spectrum this chain offers when used in a $^{222}$Rn-loaded scintillator source can be seen in Figure~\ref{fig:Sources_old}'s inset.}
\label{fig:rn222}
\end{figure}

The usefulness and versatility of a $^{222}$Rn-loaded scintillator source in Borexino containing negligible amounts of chemical quenchers lies in the fact that it includes all types of radiation (except neutrons) over the energy range of interest (see Figure~\ref{fig:rn222}). In particular, the $^{214}$Po peak (by itself and in conjunction with the $^{218}$Po-$^{222}$Rn lower energy peak) is very useful for MonteCarlo energy response tuning, with the inclusion of regional effects when deployed at different positions around the active volume. It is also extremely useful for the accurate determination of the position reconstruction in the detector, having a point-like response in the $\alpha$ peaks, due to their large $dE/dx$ (making the spatial smear caused by their ionization to be smaller than the detector's cm-scale spatial resolution, and therefore offering this point-like response). From there, a precise determination of the effective index of refraction, $n_{eff}$, of the LS is also possible: through the determination of the true-to-reconstructed source position, systematic biases due to the frequency dispersion behavior of the scintillation wave packets can be quantified and re-tuned for by including these effects in the effective parameter. 

These issues are especially important for the 2017 calibration campaign, which will be more focused towards the SOX program, where most of the statistics will be gathered at peripheral areas of the IV, in its bottom hemisphere. For those reasons, $^{222}$Rn source deployment is seen as an extremely important part of this new calibration campaign, only in second place after a neutron source (which would mimic the delayed signal from an antineutrino capture).

The mixture utilized needs to match as well as possible the scintillator that would surround it when deployed in the IV, both regarding the transparency of the mixture, its light-yield and scintillation characteristics (related directly to the quencher concentration the source's scintillator contained, see Section~\ref{subsec:quenching}), as well as avoiding too large bubbles or other optical path effects that could interfere with the scintillation light propagation. Moreover, it must ensure no long-term radiocontaminants are left inside the detector (i.e. easy thorough cleaning and low intrinsic backgrounds on its external surfaces). Its activity must also be high enough to allow for adequate statistics collection. Additionally, a long enough life with reasonably high trigger levels before it decays away must be guaranteed --while making it low enough that it would not cause saturation or instability problems with the DAQ trigger. This was limited to $\mathcal{O}$(100) Hz with the old Borexino Trigger Board (BTB), requirement which is expected to be significantly alleviated with the new trigger installed in the summer of 2016.

In conclusion, there are three main objectives to be met with the technique described in the next sections: (i) ensure external radiocleanliness to avoid insertion of long-lasting radioisotopes into the extremely radiopure LS inside Borexino's IV; (ii) maximize source activity (within the operational limits that constrain it to  $\mathcal{O}$(100) Bq at deployment) with a procedure as simple and reproducible as possible; and (iii) minimize the introduction of chemical quencher agents in the LS throughout this procedure, to keep the LS' properties as close as possible to the original ones found in Borexino's active volume.

\section{Technique}
\label{sec:technique}

\subsection{Precedents}
\label{subsec:precedents}

Three sources based on $^{222}$Rn-loaded PC+PPO scintillator were used extensively during the 2008-10 first Borexino calibration campaign\cite{calib}\cite{Hardy}, along with a $^{222}$Rn-deposited low-pressure vial used without scintillator. During the 2017 calibration campaign in preparation for the SOX program, $^{222}$Rn LS-based sources with as little chemical quencher concentration as possible will also be extensively used.
 
\begin{figure}[ht]
\centering\small\includegraphics[width=0.3\linewidth]{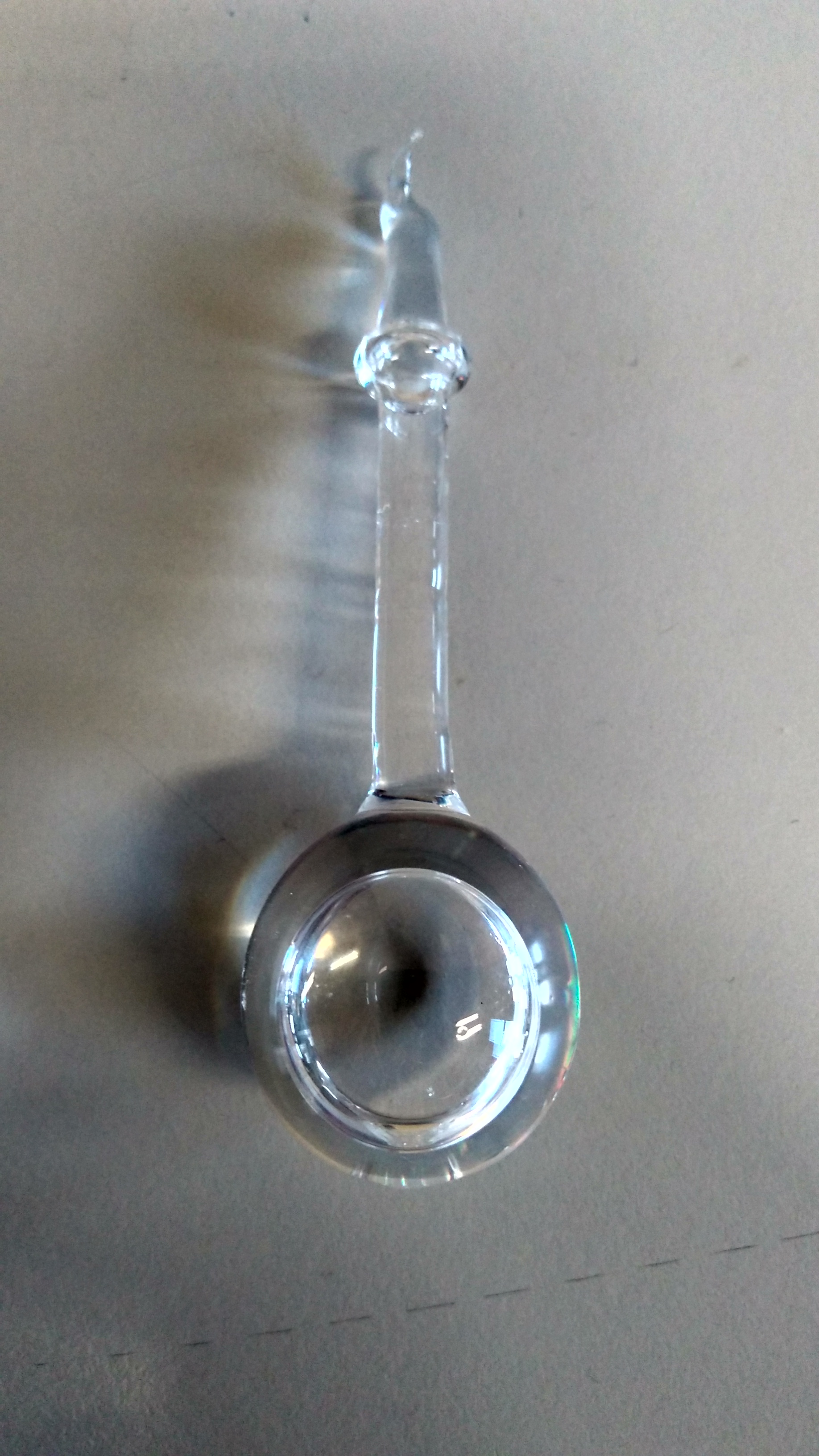}
\caption{Sealed sample source vial, featuring the $\sim$1''-diameter quartz sphere filled with LS (where the $^{222}$Rn would be dissolved), the connection Pyrex neck, joined to the sphere with a quartz-to-glass transition, and the protrusion ring for safe attachment to the Source Deployment System of Borexino. The top of the vial has been fire-sealed at the mid-point of the glass constriction, discarding the sacrificial neck above it.}
\label{fig:vial}
\end{figure}

The radon source containment ampoules are the standard Borexino calibration quartz vials\cite{Hardy}, used also for the $\gamma$ sources. These vials are intended to have a very simple design, in order to reduce failure modes, yet fulfill three stringent requirements: \textit{(i)} secure attachment to the source deployment mechanism, \textit{(ii)} allow for thorough surface cleaning with minimal hard-to-reach or non-smooth surfaces, in order to preserve to the highest extent practicable the unprecedented radiopurity of Borexino's active volume scintillator fluid, and \textit{(iii)} transmit ultraviolet light throughout as much of the PPO fluor spectrum as possible. A Pyrex glass neck is joined to an opening on the top of this 1''-diameter quartz sphere to allow for liquid and gas loading (see Figure~\ref{fig:vial}). The 0.19''-diameter glass neck features a constriction at $\sim$3.4 cm from the sphere-to-neck joint, designed to ease fire-sealing of the system once filled with the Rn-loaded scintillator, as well as a small protrusion around the neck to ensure no slippage can occur when secured to the detector's source holder. A sacrificial neck length above the constriction is used to attach the vial to the fluid loading station described in Section~\ref{subsec:newsystem}. Vials are thoroughly cleaned in its interior with several baths of acetone, isopropanol, critical-cleaning detergent and de-ionized water.

The technique for creating this kind of sources was first developed and used operationally during the 2007-09 internal calibration period. The characteristics of the sources produced during that development are reported in \cite{Hardy}. Most of them were mixed with toluene in order to include a certain amount of $^{14}$C activity (50 $\mu$L or $\sim$30 Bq), useful to characterize detection response in Borexino's lowest energy threshold. This addition is no longer performed, as $^{14}$C has been well understood since then\cite{pp}, and maximizing the $^{222}$Rn activity calls for an increase in the trigger threshold, in order to keep out the $\sim$30 Hz of scintillator-contained $^{14}$C present in Borexino's signal by default.

The key aspect that differentiates the current technique from previous attempts at high-activity liquid scintillator $^{222}$Rn loading is the deposition of the $^{222}$Rn in solid form on the inner vial walls, through freezing by vial immersion in liquid nitrogen (boiling point at normal pressure and temperature of 77.36 K). In particular, radon solidifies at 202 K. This technique does not rely on radon being dissolved inside the scintillator (and staying in it while more is being loaded) like previous works, and therefore allows for much higher resulting activities. The radon-loaded UHPN whose carefully-controlled flow is directed through the $^{222}$Rn generator is directed through a retractable thin metal capillary to make it impinge on the internal vial walls while immersed in the liquid nitrogen bath. This was shown to provide an effective retention mechanism for part of the $^{222}$Rn mixed in the nitrogen flow, while allowing for the latter to escape through an exhaust port that closes the circuit. The microscopic deposition mechanism for the $^{222}$Rn has not been researched in detail for this work (see \cite{rn_fcc} for radon crystal growth), but in principle should enable to fully predict deposition rates depending on the flux of $^{222}$Rn-loaded gas entering the freezing zone, which clearly will not be linear (depending on nucleation sites, which could na\"ively be assumed to grow as $\sim r^2$ assuming a small nucleation region would preferentially grow radially along the deposition surface). However, as will be shown henceforth, reasonably consistent loading rates have been achieved. A detailed sequential description of the refined technique can be found in Section~\ref{subsec:newsystem}.

\subsection{Quenching causes and effects on signal}
\label{subsec:quenching}

Liquid scintillator detectors such as Borexino depend on the scintillation process, where fluorescence photons are emitted when charged incoming particles lose energy either by ionizing or exciting the solvent (PC). This is a process dependent on the nature of the charged particle itself, its energy and the solvent-fluor interactions (PC-PPO). The time distribution and the light yield of the emitted fluorescence photons depend on the various molecular processes taking place as a consequence of the particle energy loss. They do not only depend on the energy deposit in the scintillator but mainly on how the energy is released, \textit{i.e.} on the value of $dE/dx$ and on the type of incident particle. Heavy ionizing particles like $\alpha$s feature a large $dE/dx$, and produce large ionization or excitation density, thereby increasing the probability to get the triplet excitation state $T_{10}$ --and consequently, delayed fluorescence. Such large ionization or excitation densities favor molecular processes in which the energy is dissipated in non-radiative ways, which results in the quenching of the scintillation light. This is referred to as ionization quenching\cite{Birks}. Proper understanding and modeling of this effect is crucial for detector operations\cite{MC}.

Scintillator quenching effects similar to ionization quenching, but affecting all incident radiations exciting the scintillator, can also be produced by contaminating substances which absorb or modify the scintillation light before it reaches the fluor. This has the net effect of reducing the light yield and, therefore, modifying the energy response of the scintillator. Effectively, a shifting and scaling of the energy scale will be observed. Therefore, if a LS-based calibration source contains quenchers in its scintillator, its usefulness for proper energy reconstruction and light yield determination (especially for short-range radiation) will be much diminished. In other words, its properties will differ from those of the detector's scintillator around it, whose response is the subject of the calibration.

In the case of PC, exposure to oxygen is the most likely quenching factor. Indeed, just a short exposure to the atmosphere will severely quench its response (not counting the introduction of backgrounds dangerous to Borexino's target signal, such as $^{85}$Kr). Acute, low-level oxygen exposures however, can be mitigated by sparging (that is, areating with dry nitrogen or other inert pure gas) the scintillator to bring it back to its original chemically-unquenched state.

However, while dissolved oxygen is a known quenching agent by itself, if it is left in contact with pseudocumene for a sufficient period of time, its molecules will also react to form dimethylbenzaldehyde (DMBA) or other oxygen-containing molecules --these are chemical changes and are not removed by sparging (conveniently, DMBA has a higher boiling point than pseudocumene, thus, distillation can separate the two chemicals --although this is mostly useful for large scale scintillator treatment, such as through Borexino's purification plants\cite{purif}, and not so much for small samples such as the ones used for the $^{222}$Rn-loaded sources).

\subsection{Latest developments and technique}
\label{subsec:newsystem}

\begin{figure}[tb]
\centering 
\includegraphics[width=0.8\columnwidth]{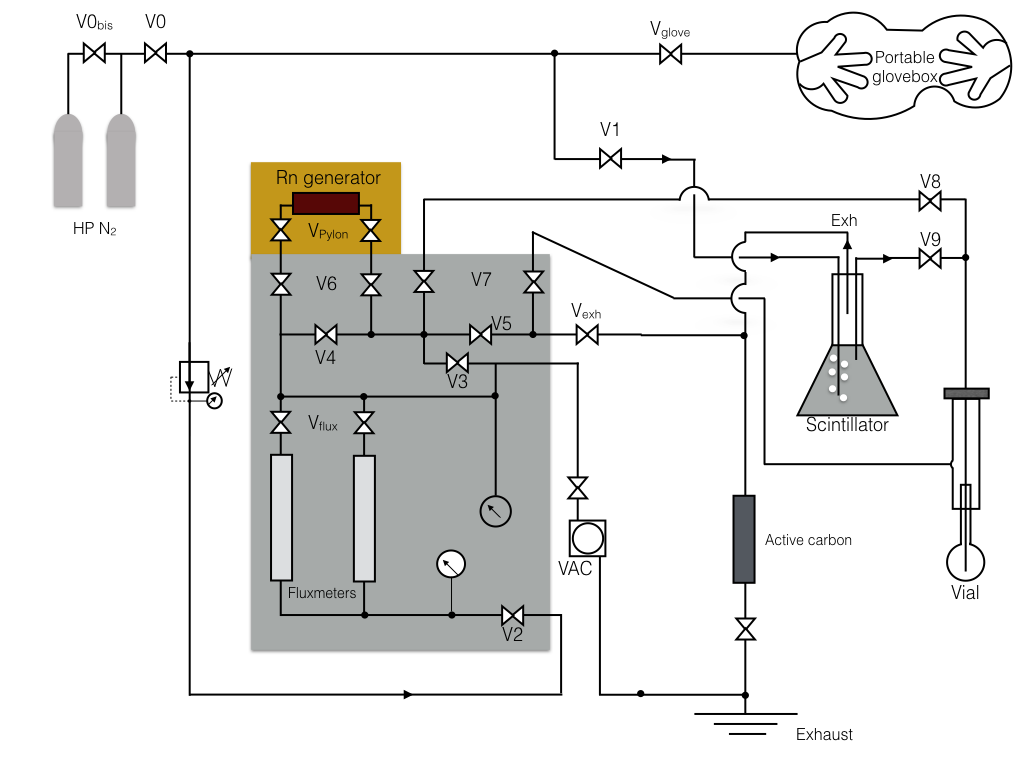} 
\caption{Radon loading setup diagram for the high-activity $^{222}$Rn sources with negligible amounts of quenchers developed in the Federico II University of Naples. The upper left cylinders represent the High Purity Nitrogen supply. The portable glovebox was used for (dis)assembly of pieces, such as the vial or scintillator flask, when exposure to atmospheric oxygen is not desired. The grey box represents the monolithic panel employed for pressure determination and core valve control, scavenged from CTF's source production campaigns. The activated carbon filter is employed to avoid indirect transport of PC vapors to the rotary vacuum pump when the upstream panel lines are evacuated, even if $V_{exh}$ is nominally closed.}
\label{fig:Rn_setup} 
\end{figure}

\begin{figure}[tb]
\centering 
\includegraphics[width=0.8\columnwidth]{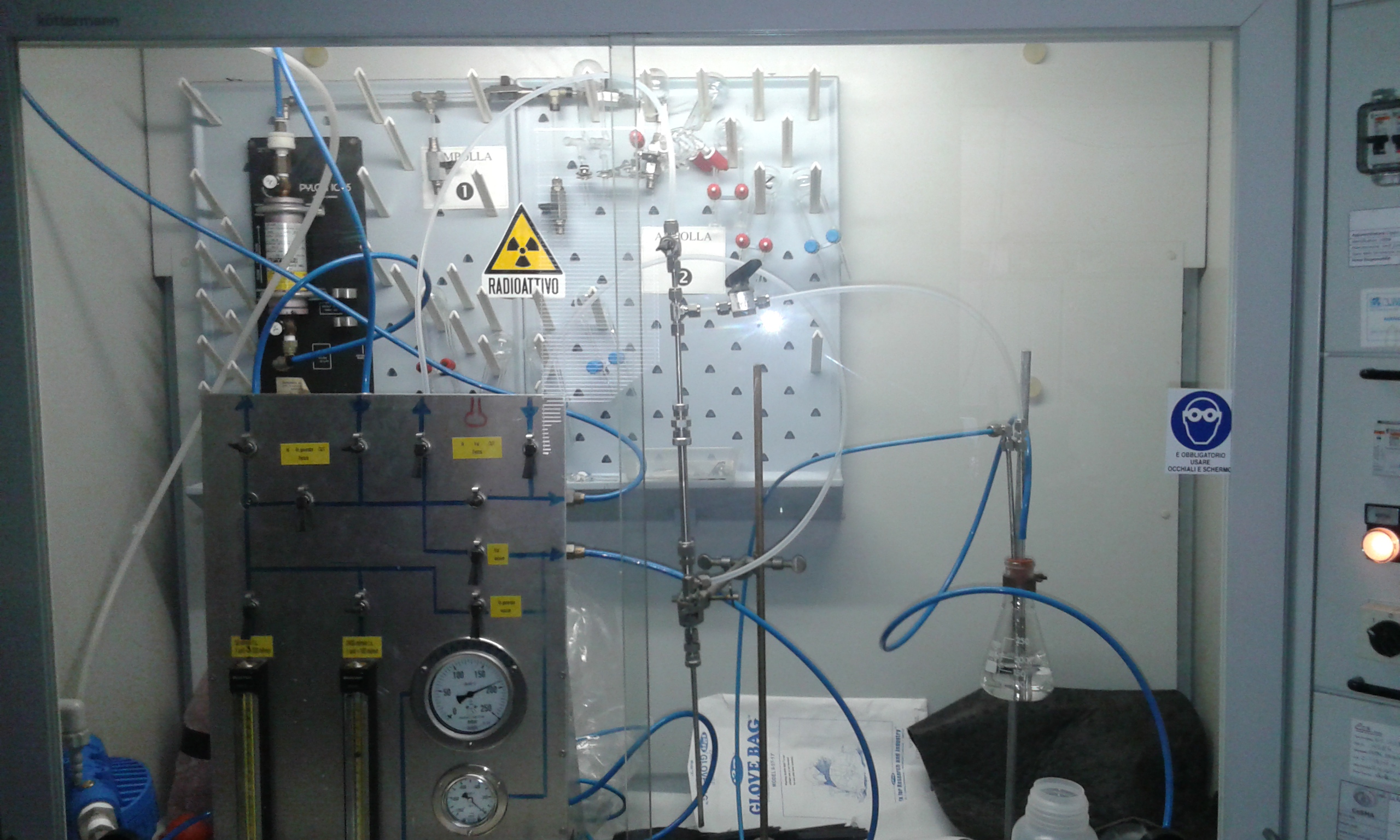} 
\caption{Picture of the final version of the radon loading station.}
\label{fig:Rn_setup_pic} 
\end{figure}

The loading station was completely re-designed from the 2007-09 system, and the few similarities shared with the previous setup were due to convergent designs aiming for similar objectives. However, the current aim was to lower the quenching to the minimum, which was not reliably achieved in an operational manner during the first calibration campaign\cite{Hardy}. Furthermore, the source loading had to be demonstrated to be scalable with reliability to achieve the activity required. A diagram of the system is shown in Figure~\ref{fig:Rn_setup}, and a photograph of it located in the fume hood in INFN Naples' laboratory in University Federico II is shown in Figure~\ref{fig:Rn_setup_pic}.

The system consists of a heritage panel from the Counting Test Facility (CTF)'s loading experiments, containing the core fittings and instrumentation for gas distribution. It was kept unmodified, except for a thorough interior cleaning, since the Swagelock fittings and welds had already been proven and leak-checked. This panel includes two parallel flux-meters (coarse and fine) for flow control, a manometer in series with those for pressure monitoring, and the radon generator in/out ports, as well as those for the source loading. Finally, an exhaust port for the source was also provided, as well as a vacuum port for fluid drawing, also useful for line evacuation, purging and cleaning.

The vacuum pump employed was a low-power membrane pump capable of bringing line pressure down to a few millibar, more than enough for source creation purposes, and adequate after a few iterations for line evacuation and pumping to avoid oxygen presence.

An ultra-high-purity nitrogen (UHPN$_2$) supply was connected to three different points in the system: 

\begin{enumerate}
\item The flow intake ($V2$) in the panel for the drawing of gas from the radon generator, as well as for the flow supply to the source vial.
\item The scintillator flask containing the liquid used for source filling, that was kept under constant, low-intensity sparging as a precaution against inadvertent exposure to oxygen or other gaseous quenching agents\footnote{Exhaust for this flask was directed through an activated carbon filter to the fume hood ventilation system. While the fume hood wouldn't necessitate this precaution, the connection to the panel through the $V_{exh}$-to-$V5$ line meant that, in the absence of the filter, PC condensation would have the chance to reach the vacuum pump.}.
\item A portable glovebox that served as an oxygen-free environment for flask filling operations, vial adjustment when filled with scintillator, and other quenching-critical operations.
\end{enumerate}

The flask containing the scintillator (drawn directly from Borexino's IV under LAKN atmosphere in CR4's facilities) was connected to the system through a triply-perforated rubber cap that ensured a reasonably hermetic closure. The overpressure inside the flask caused by the UHPN$_2$ bubbling was kept in a safe range to avoid it forcing the cap out or damaging the flask, as well as to prevent too much scintillator sloshing and bubble ingestion during scintillator drawing to the vial. The two connecting tubes going inside the flask and in contact with the scintillator (sparging line and drawing line) are Pyrex glass, while the exhaust tube is Teflon.

All the lines not contained within the panel are Teflon tubes with steel or plastic (depending on the criticality of their position within the system) Rapid Fittings, except for gas-only lines which are polyethylene tubing.

The source vial is seamlessly held in a sleeve-needle leak-tight holder since the beginning of the procedure until sealing. It is composed of a thin steel tube through which the radon-loaded nitrogen flux and drawn scintillator is directed into the vial; and a concentric, larger diameter sleeve that serves as an exhaust container for the nitrogen flux. This sleeve features two viton O-ring fixtures at its top and bottom that serve as the needle height regulator and vial holder, respectively: the top fixture can be loosened enough to permit the needle to be retracted before or inserted beyond the fire-sealing neck constriction, while keeping the overpressure inside it to make sure no quenching agents get into the system.

\begin{figure}[tb]
\centering 
\includegraphics[width=0.6\columnwidth]{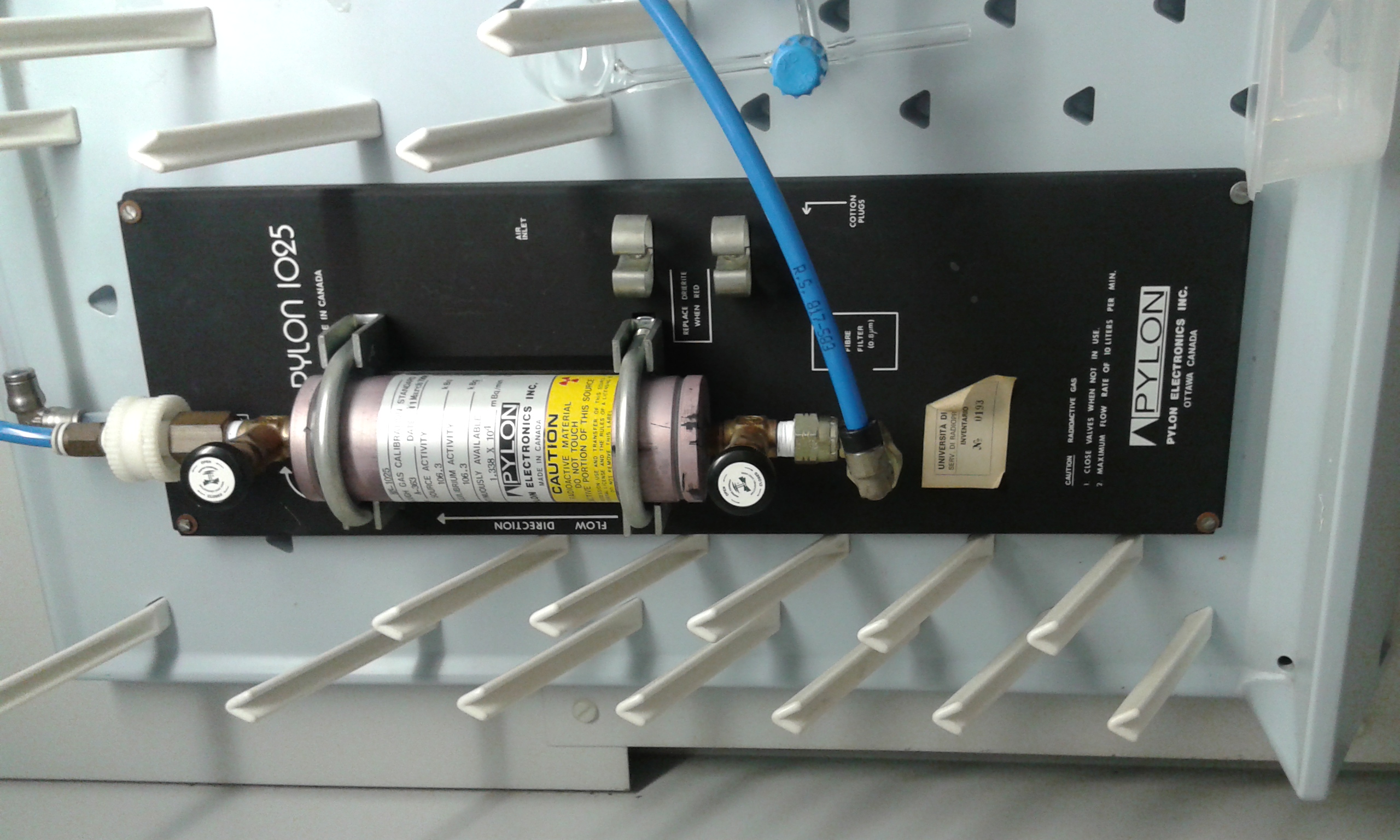} 
\caption{Pylon RN-1025 flowthrough gas source.}
\label{fig:Pylon} 
\end{figure}

\begin{figure}[tb]
\centering 
\includegraphics[width=0.7\columnwidth]{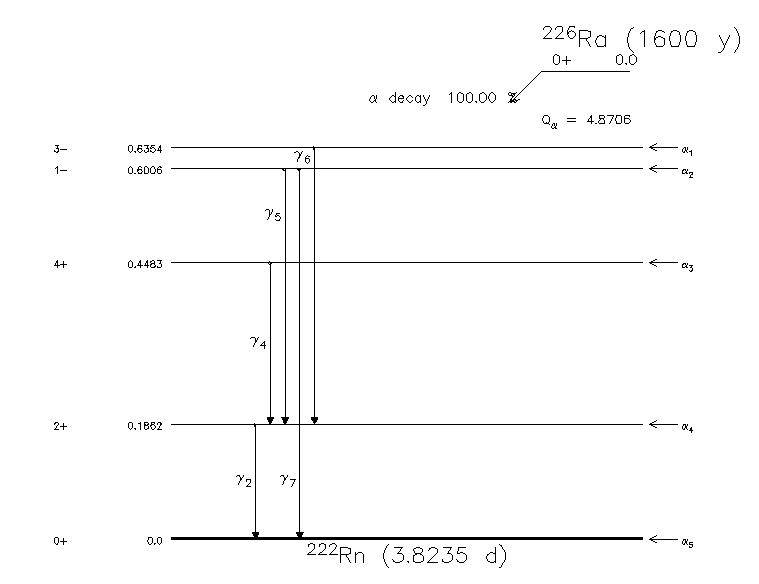} 
\caption{Radium-226 decay scheme, yielding the $^{222}$Rn needed for the calibration sources through the 100$\%$ branching ratio $\alpha$ decay, as well as the possible de-excitation $\gamma$s. The $\sim$1600 year half-life means that the radon generator emanation activity will be constant for any Borexino-related program.}
\label{fig:226ra} 
\end{figure}

The radon generator is a commercially-available Pylon RN-1025 flowthrough source\cite{Pylon}. This source is an aluminium cylinder with two attach fittings (see Figure~\ref{fig:Pylon}), with a capsule containing $^{226}$Ra salts (see decay scheme in Figure~\ref{fig:226ra}), sandwiched between particulate filters to avoid release of non-gaseous substances. It has an equilibrium activity of 106$^{+25}_{-10}$kBq, with a rated stable emanation of 13.4 Bq/min under continuous gas flux (maximum flow rate: 10 L/min).

After assembly, the system was air- and vacuum-cleaned, and several sacrificial PC drawings were performed through the tubing areas where scintillator was expected to flow through. In spite of those precautions, particulate contamination was still present in the first few test runs. For that reason, and since a system improvement was made after high-activity loading feasibility runs were completed, in order to reduce oxygen levels to $<$ppm levels, the system was thoroughly re-cleaned again, including flushing a hot Cytranox detergent mixture through the piping that would see liquid flow. This was repeatedly rinsed with water flows and airflow-vacuum cycles afterward. However, the first few subsequent test drawings showed foaming and a degree of white particulates that, while not impeding the successful demonstration of high-activity sources with negligible amounts of chemical quenchers, showed a potentially undesirable feature for future operational sources. Posterior isopropanol and PC flushings showed their effectiveness through the absence of noticeable contamination when performing later trial source creation runs.

\section{Results}
\label{sec:results}

The main results obtained both during the technique development in the 2000s and the latest results with the improved technique in recent years are reported here. In particular, the main goals listed at the end of Section~\ref{subsec:222rn} and their technical solutions detailed here are: (i) external radiocleanliness is ensured by the vial's design; (ii) maximal activity is achieved thanks to the radon-freezing process by which liquid nitrogen temperatures are kept at the vial walls in order for the $^{222}$Rn to solidify on their interior side while allowing for the radon-loaded UHPN flux to continue for as long as necessary; and (iii) minimal quencher concentration with respect to the initial LS used for filling the source vial is reached thanks to the extreme cleanliness and leak-tightness implemented for the otherwise relatively simple-to-operate setup, as well as the use of ultra high-purity nitrogen in all phases where LS may be present.

\subsection{Old sources data}
\label{subsec:oldsources_data}

Sources created in 2008-10 were characterized by including the $^{14}$C-containing toluene solution ($\sim$100 $\mu$L, or $\sim$30 Bq), inserted in the vial before radon loading and scintillator withdrawal. Activity determination was not deemed extremely important for this level of development, compared to relative quenching results, and only single p.e. double-pulse spectrometers were employed. For this reason, the numbers reported in Table~\ref{table:results}'s "VT" (Virginia Tech, since they were fabricated there) section are understood to be mostly indicative. Further, an absolute energy calibration of this detector was not needed since it would be determined when the sources were introduced in Borexino. Therefore, only relative quenching was determined.

\begin{figure}[ht]
\centering\small\includegraphics[width=1\linewidth]{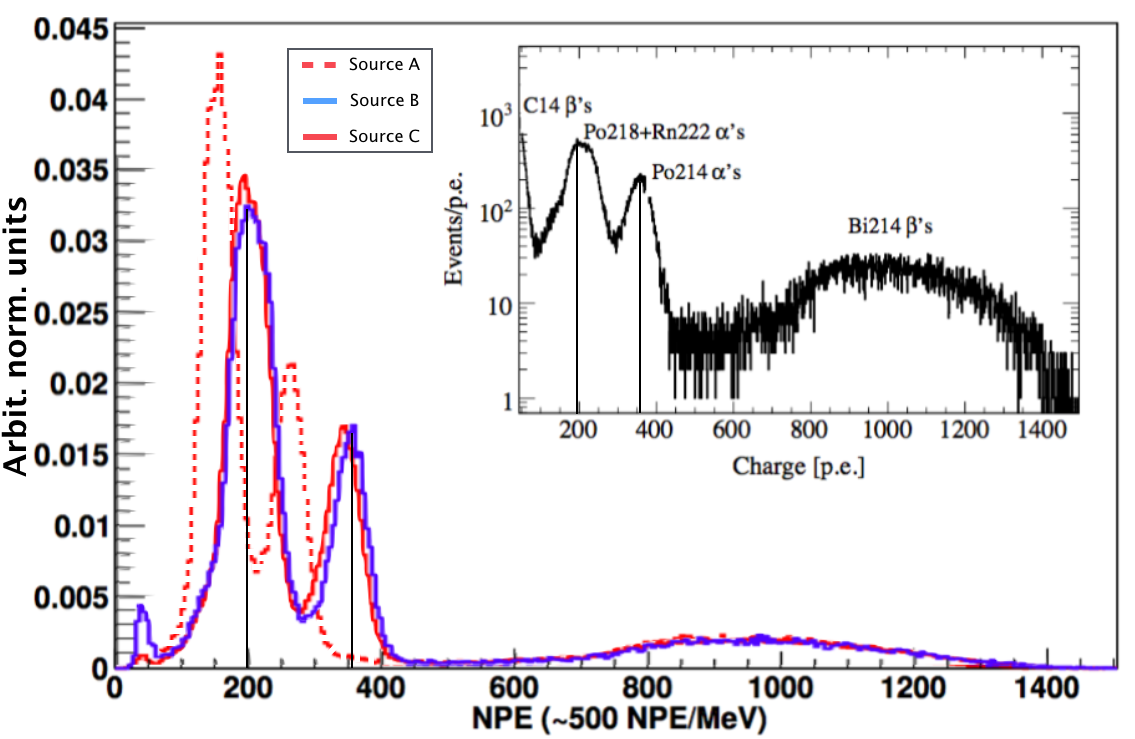}
\caption{Results from the first calibration campaing's sources, used in 2008-2009, from Borexino data taken at deployment, and a fiducial cut of $<$50 cm from the detector's center. These sources, apart from $^{222}$Rn loading, also included extra $^{14}$C. An evident quencher presence can be seen in the red curves, used in the source A for the on-axis (October 2008, dashed line) deploy and source C, used on the second off-axis deploy (June 2009, continuous line), estimated at $\sim$30$\%$ and $\sim$5$\%$, respectively. The blue curve corresponds to the first off-axis deploy (January 2009, source B), which was the least chemically-quenched source employed in the 2007-09 calibration campaign. In the inset\cite{calib}, an annotated reference spectrum of the spectral shape expected from $^{222}$Rn in equilibrium, with three distinct peaks from different $\beta$ and (ionization-quenched) $\alpha$ decays in its chain. The horizontal displacement of the peaks in sources A and C is evident, while for B it is much less pronounced, indicating the lower quencher concentration reached in that particular source (compatible with zero, at the $<1\%$ level). Arbitrary units (Y axis) stands for the number of events in Borexino when the sources were deployed. NPE stands for Number of PhotoElectrons and is a measure of the charge collected by Borexino's PMTs --its equivalence in MeV is indicated in parenthesis.}
\label{fig:Sources_old}
\end{figure}

\begin{figure}[ht]
\centering\small\includegraphics[width=1\linewidth]{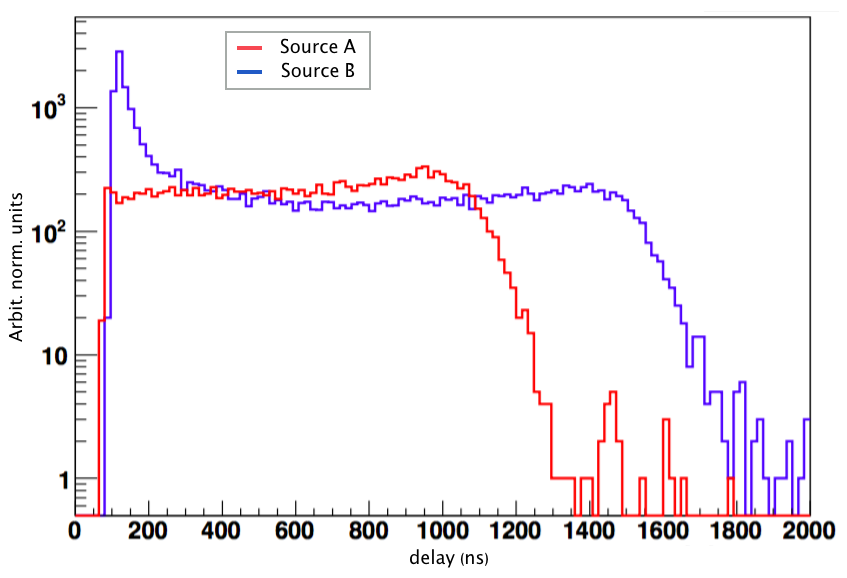}
\caption{Single photo-electron curve of the scintillator used for the first calibration campaign's on-axis sources (red, source A) and the one used for the first off-axis calibrations of January 2009 (blue, source B), shown to be the only mostly unquenched ($<1\%$) one in Figure~\ref{fig:Sources_old}. These curves show the cumulative scintillation time-decay curve for the first acquired p.e. after each triggered event\cite{Paolo_quench}. A left-shifted curve means the delayed response tail is inhibited, because of the reduced light yield the quenching agents cause, as explained also in the next section. The low-energy peak in the blue curve is $^{14}$C, quenched off the threshold in the on-axis source A.}
\label{fig:scintillator_quenching}
\end{figure}

Three operational sources (A, B and C) were created, as well as a "weak" source of $\sim$10 Bq that was just used as a comparison for the first operational source. An absolute quenching of $\sim$30$\%$ was discovered on the first source A (deployed on the detector's vertical axis, or "on-axis" for short), as well as on the "weak" reference source. This is attributed to the fact that the scintillator used was made separately from Borexino's and didn't go through the same processing. The second off-axis source C (deployed away from the vertical axis of the detector) also exhibited an amount of quenching, although somewhat smaller (estimated $\sim$5$\%$), while the intermediate source B, the first used off-axis, exhibited a negligible amount of quenching, owing to the new batch of scintillator used, carefully withdrawn from Borexino's IV to understand the impurity quenching on the first sources, as well as renewed procedures (see Figures~\ref{fig:Sources_old} and~\ref{fig:scintillator_quenching}).

In spite of the irregular, non-negligible quencher levels attained for these initial operational sources using the $^{222}$Rn LS-loading procedure, they demonstrated the technique held potential to create quencher-free sources, and gave a strong motivation for the latest technique re-design described below. Moreover, their use during the 2007-09 calibrations highlighted the importance of $^{222}$Rn sources for the understanding and tuning of Borexino's response, even if their scintillator was noticeably quenched\cite{calib}.

For a fully detailed account of the processes, setup and characterization results, refer to \cite{Hardy}.

\subsection{Newly attained sources}
\label{subsec:newsources_data}

Several test runs to verify the integrity of the new core system and the feasibility of high-intensity radon loading were performed with a simplified system (which didn't guarantee ppm oxygen removal) prior to the final assembly of the system described in the previous section. The objectives for these dry and wet dress rehearsals were:

\begin{enumerate}
\item Verify leak tightness and proper operation of the core panel.
\item Test new needle/sleeve vial holder.
\item Design and test scintillator withdrawal and sparging system.
\item Demonstrate large amounts of radon could be deposited with the liquid nitrogen bath with the current system, and adjust flushing durations to achieve the desired activity levels: dry runs (no scintillator; 164$\pm$8 Bq deposited).
\item Develop, test and refine scintillator withdrawal and vial filling procedures.
\item Demonstrate reliable line clearing techniques to avoid vial overfill after the initial filling operation was completed.
\item Test acceptable limit for vial ampoule filling.
\item Practice integrated (loading + filling + freezing) operations: wet runs.
\item Develop and practice source fire-sealing. Previous sources had relied on professional glass-makers who were not available on this occasion.
\item Practice legacy procedures.
\end{enumerate}

Several test vials were employed, some of which could be re-used since no loading or fire-sealing had taken place. By the end of this phase, all objectives were achieved except the 6th point above: reliable line clearing techniques after the primary filling operation was complete. Reproducibility of the fill level was poor and accidental overfilling was very likely, although its amount was very difficult to predict, so it could not be reliably estimated and corrected for during the primary filling. Additionally, as was expected, scintillator quenching was severe.

Regarding activity levels, the radon laboratory in the Federico II University of Naples provided a very sensitive and well-calibrated Ortec hyperpure germanium crystal spectrometer (gamma-X type\cite{germanium}), allowing to very well characterize the fabricated test (and operational) sources in-situ, mere seconds after their sealing. This detector features a beryllium entrance window for the emitted radiation, allowing to bring the observed spectra's lower limit to $\sim$20 keV, with a 2 keV energy resolution (at 1.33 MeV) and a relative efficiency of around 48$\%$. Furthermore, the Ortec Gammavision v.6.0 software\cite{Ortec} provided a quick and precise assessment for the equilibrium activity measurements. This activity is determined from the $\gamma$-emitting daughters from the $^{222}$Rn decay, in particular $^{214}$Pb and $^{214}$Bi, which is then assumed to come from the radon $\alpha$ decay. TThis condition is verified after 3 hours following sealing of the radon source. The $\gamma$ lines are Gaussian-fitted by the software to yield the activity and uncertainty.

A test source was first loaded. It was intended to be left loading for 36 minutes at a measured UHPN$_2$ flux of 21 mL/min. However, because of problems with circuit closure, only partial deposition had been happening for most of this time --but conversely got a "hit" of the $^{222}$Rn that had accumulated in the RN-1025 generator once it was properly connected. Final activity was 363$\pm$16 Bq. Considering as a measure of the efficiency of the loading procedure the specific activity per unit of time and flux, it would yield $\sim$0.48$\pm$0.02 Bq/(min$\cdot$mL), but as mentioned, the deposition in this test source was not optimal for the technical reasons described above.

Subsequently, the fluid-handling part of the system was completely renewed, except for the scintillator flask and the needle/sleeve holder, by using the new high-quality metal Rapid Fittings and relying on more extensive use of thoroughly cleaned Teflon tubing. Procedurally, routine high-fluence UHPN$_2$ flushing and vacuum pumping cycles were put in place to evacuate the system of oxygen to the $\sim$ppm level. A more detailed overview of the procedure can be found in \cite{my_thesis}.

\begin{figure}[tb]
\centering 
\includegraphics[width=0.7\columnwidth]{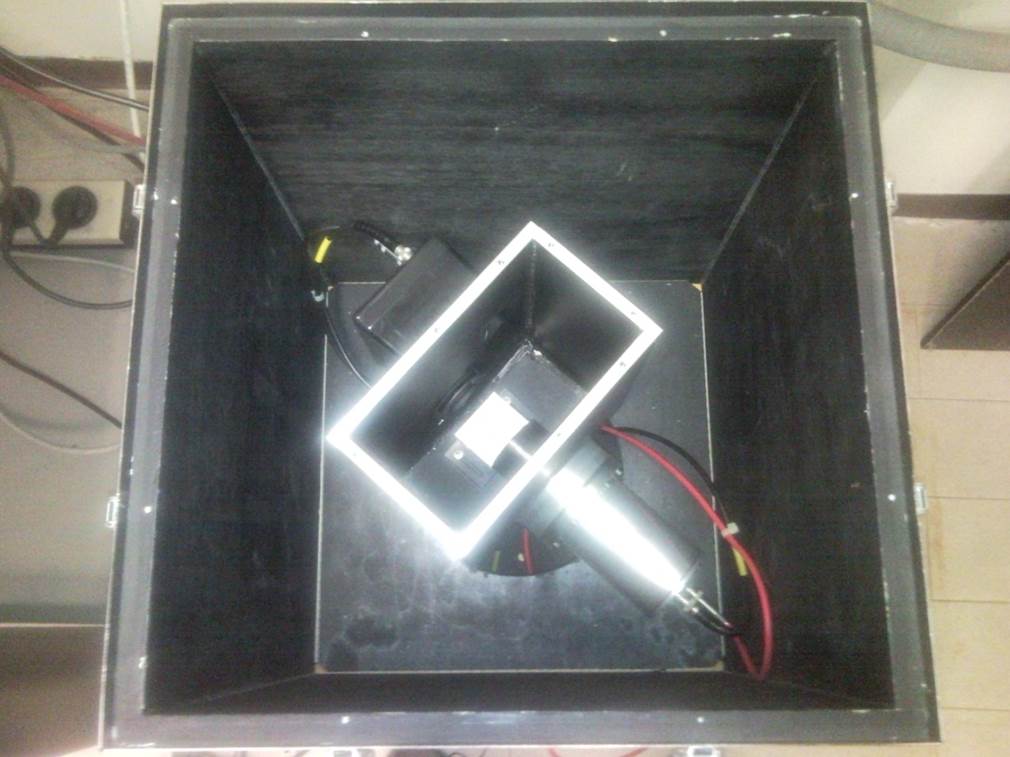} 
\caption{Photograph of the single-photon setup used for the source's scintillation decay time measurements shown in Figure~\ref{fig:Sources_all}. A detailed diagram of the technical details of the setup can be found in \cite{Paolo_quench}.}
\label{fig:Paolo_PMT} 
\end{figure}

\begin{figure}[ht]
\centering\small\includegraphics[width=1\linewidth]{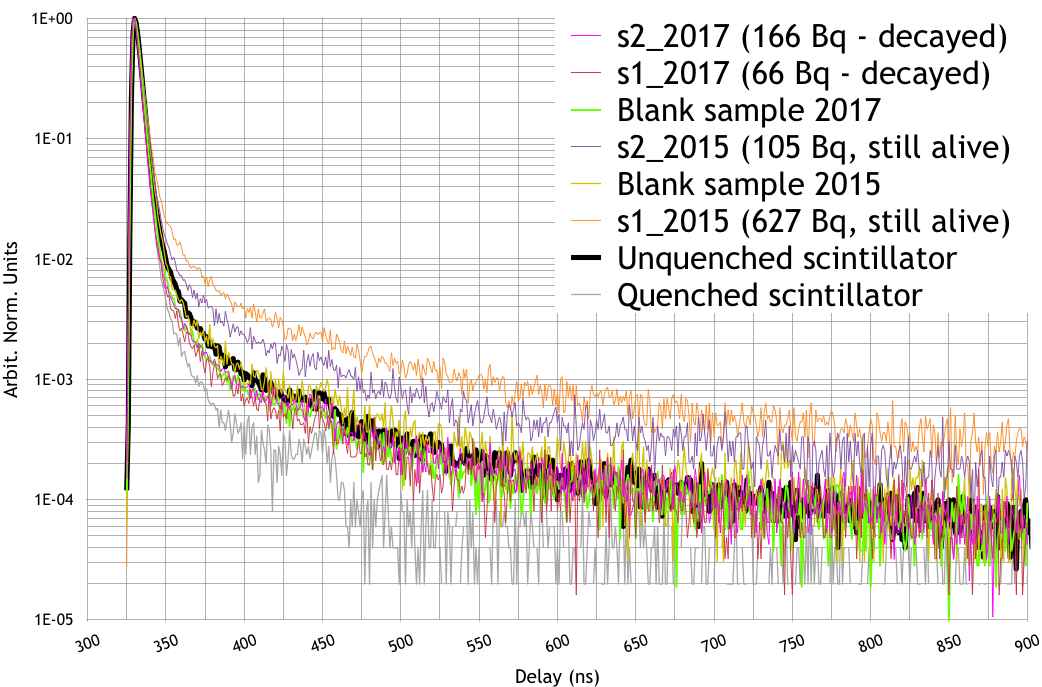}
\caption{Single photo-electron curve from all sources (and blanks, meaning non-loaded scintillator that passed through the same procedure and lines as the one used for a real source) fabricated with the new, improved setup at INFN Naples. For reference, curves from a scintillator (PC+PPO) mixture drawn directly from Borexino's IV, considered to contain negligible quencher amounts, and the first sample source (containing quenchers) are also included. Note the still-alive sources (those in 2015) show a higher-than-baseline response because of the extra hits coming from the $^{222}$Rn decays. The "delay" label in the abscises axis indicates the single p.e. delay in the signal used for the scintillator quenching measurement as per the technique indicated in~\cite{Paolo_quench}. Note the similarity between the S1/S2\_2017 curves and the 2017 blank, indicating that although the scintillator batch exhibited some quencher content, the loading process did not significantly worsen it, as detailed also in Table~\ref{table:results}.}
\label{fig:Sources_all}
\end{figure}

Results showing the chemical quenching of the sources were provided by Milan's University time decay profile measurement setup for scintillator mixtures\cite{Paolo_quench} (see Figure~\ref{fig:Paolo_PMT}), based on a previous heritage design\cite{Gioacchino_PMT}, featuring a weak $^{60}$Co excitation source and two photomultiplier tubes: a strongly-coupled one (high-level PMT) and a loosely-coupled one (low-level or fluorescence PMT, for single-p.e. sampling) providing the stop and start signals, respectively. The loose coupling between the specimen and the fluorescence PMT is achieved with a set of neutral filters. An electronic DAq system consisting of a counter, constant fraction discriminators connected to the anode, timer and coincidence units and a digitizer (10 bit, 2 Gb/s Agilent Technology) was integrated through a LabView software architecture\cite{LabView}. Particular attention was devoted to the long scintillation decay time profile tail, since it is extremely sensitive to quenching, both by shortening of the long scintillation response in time as well as reducing its yield. Additionally, a single-PMT setup was used for the 2017 sources in order to double-check the germanium counter results, which yielded positive results --and also confirmed the decay peaks, as well as the $^{60}$Co source used for excitation, were not shifted relative to each other, reaffirming the absence of relative quenching. The scintillation curves may be seen in Figure~\ref{fig:Sources_all}, where responses below the reference curve can be interpreted as having an amount of quenchers present, while those coincident with it (or above it, if they additionally have some radioactive isotopes still alive in them) show negligible chemical quenching. The new loading system was thus shown to have provided chemical quenching results compatible with pure scintillator drawn from Borexino's IV (which was also used as a control reference baseline assumed to contain negligible amounts of quenchers). 

The first loaded source (S1\_2015) spent about an hour (62 minutes) in loading mode, with a UHPN$_2$ flux of 40 mL/min, and was measured to have $\sim$650 Bq just after sealing. An overnight measurement (12 h of integration time) led to the more precise determination of 665$\pm$28 Bq. This would yield a loading efficiency of 0.269$\pm$0.012 Bq/(min$\cdot$mL).

A second source (S2\_2015) with 20 minutes of loading time and the same flux yielded 105$\pm$5 Bq, or 0.131$\pm$0.001 Bq/(min$\cdot$mL). 

In light of the preparation of these sources, as well as the dry runs performed with the old system, it is apparent a high level of $^{222}$Rn deposition is achievable and repeatable while providing an environment for the scintillator compatible with negligible amounts of quenching agents. Actual demonstrated initial source activities are far beyond the requirement of $\sim$120 Bq at production, up to more than 5 times. Loading flow time can be extended if necessary, and is in principle not constrained by any factor other than fluid (UHPN$_2$ and liquid nitrogen) availability. It is then expected larger activities can easily be achieved with the same setup, raising the initial activity ceiling to no less than 1 kBq, which could potentially allow for more flexibility in the calibration campaign, either by increasing the available wait time to $\sim$10 days between production and deployment in Borexino, or by shortening data-taking time. 

Further, an approximate estimate of $\sim$0.2 Bq/(min$\cdot$mL$_{UHPN_2}$) of average deposition efficiency is expectable from the produced sources, and can conceivably be brought up to twice as much by not performing a pre-flush of the radon generator, thereby profiting from the amount accumulated during its previous dormant phase. This option, however, is understood to not be without risks, since air leakage into the RN-1025 might introduce quenching agents that could negatively affect the source's light yield and hence its usefulness as a calibration tool.

Finally, two further sources (S1\_2017 and S2\_2017; along with a blank non-loaded sample) were completed in May 2017, in order to fully certify the new process and setup for the approaching calibration campaign in late 2017. These sources, as was later discovered, used a scintillator batch that was pre-extracted from Borexino before the sample used for the 2015 sources was withdrawn. This meant it was expected to be less pure (both because it was withdrawn under normal atmosphere --and thus exposed to oxygen, without being sparged immediately afterwards, and also because it was expected to have more impurities), although great effort was put into trying to thoroughly sparge it before the 2017 tests. However, as can be seen in Figure~\ref{fig:Sources_all}, the baseline quenching level for this scintillator was slightly larger than for the older ones, without arriving to the large quenching levels seen in 2008. It is apparent too, however, that relative quenching was not affected by the loading procedure. 

\begin{table*}[t]
\centering
\begin{tabular}{l r r r r}
\toprule
 & \textbf{Activity} & \textbf{Deposition rate} & \textbf{Quenching (\%)} & \textbf{Comments} \\
 & \textbf{\textit{(Bq)}} & \textit{\textbf{(Bq/(mL$\cdot$min)}} & & \\
\midrule
VT & $\sim$550$\pm$100 & ? & $>$30($\pm$5) & \footnotesize{On-axis source A, bad scintillator} \\
 & 425$\pm$20 & 0.47$\pm$0.01 (?) & $<$1 & \footnotesize{1$^{st}$ off-axis source B} \\
 & & & & \footnotesize{Newly withdrawn scintillator} \\
 & 1840$\pm$60 & 2.6$\pm$0.1 & $\sim$5($\pm$1) & \footnotesize{2$^{nd}$ off-axis source C} \\
 & & & & \footnotesize{Suspect Pylon overnight purging} \\
 & & & & \footnotesize{("hit" of accumulated Rn)} \\
\midrule
Naples & 363$\pm$16 & 0.48$\pm$0.02 & $>$30($\pm$5) & \footnotesize{Initial test source, Small "hit" of accumulated Rn} \\
 & & & & \footnotesize{probable oxygen exposure}Ê\\
 & 665$\pm$28 & 0.269$\pm$0.012 & $<$1 & S1\_2015\\
 & 105$\pm$5 & 0.131$\pm$0.001 & $<$1 & S2\_2015\\
 & 53.3$\pm$5.1 & 0.042$\pm$0.004 & $<$18 & \footnotesize{S1\_2017; Lost liquid N$_2$ bath for a few minutes,}\\
 & & & & \footnotesize{open to atmosphere while frozen,}Ê\\
 & & & & \footnotesize{less pure scintillator batch} \\
 & & & & \footnotesize{(consistent with blank)} \\
 & 166$\pm$21 & 0.10$\pm$0.01 & $<$15 & \footnotesize{S2\_2017; Less pure scintillator batch}\\
 & & & & \footnotesize{(consistent with blank)} \\
\bottomrule
\end{tabular}
\caption{Summary table of the achieved $^{222}$Rn sources both during the first calibration campaign and during the recent development of the new low-quenching system. Note the $\sim$15-18$\%$ quenching in the 2017 sources is not considered to indicate a problem with the procedure, but rather is consistent with the less pure scintillator batch used for these sources, as can be better seen in Figure~\ref{fig:Sources_all}. The extra $\sim$3$\%$ in S1\_2017 can be explained by the accidental opening to atmosphere while frozen, as indicated.}
\label{table:results}
\end{table*}

\section{Conclusions and prospects}
\label{sec:conc}

A reproducible method for the creation of $^{222}$Rn-loaded liquid-scintillator-based $\mathcal{O}$(kBq) small-sized sources with negligible amounts of quenching agents has been demonstrated. Furthermore, since its first operational use during Borexino's early-life calibration campaigns (2008-10), the technique has been replicated and improved, by a different operational group and using new hardware. The new loading setup is geographically much closer to the experiment than before, enabling more flexibility and schedule margin in utilizing them in the detector. This is of great importance to Borexino's upcoming 2017 end-of-Phase II/pre-SOX calibration campaign, expected by the end of 2017. The procedure to avoid quenching in the scintillator employed for the sources has been repeatedly and reliably demonstrated, in two separate test campaigns during which 4 sources and 2 blanks (non-radon-loaded vials whose scintillator went through the same processing as real sources) were created. Precision characterization was performed using three different setups (single- and double-PMT spectrometers, and a cryogenic germanium counter), also building upon the analyses carried out during the first phase of technique development, which only utilized a double-PMT setup (and the Borexino detector itself, upon calibration deployment). Order-of-magnitude loading efficiency consistency is reported too.

The technique will see operational use in the late-2017 SOX calibration campaign, where one or several $>$100 Bq $^{222}$Rn sources will be deployed in an estimated $\sim$200 positions, with special emphasis on mapping the position and energy reconstruction of the IV in its bottom hemisphere, and closer to the vessel limit than achieved in 2008/09 campaigns. This will also tackle other important high-precision detector characterization objectives not attempted before. We believe the technique, additonally, may have a wider use in other low-background experiments where such calibration sources could be of interest. 

\section{Acknowledgements}

The Borexino program is made possible by funding from INFN (Italy), NSF (USA), BMBF, DFG, HGF and MPG
(Germany), RFBR (Grants 16-02-01026 A, 15-02-02117 A, 16-29-13014 ofi-m, 17-02-00305 A) (Russia), and NCN
Poland (Grant No. UMO-2013/10/E/ST2/00180). The indispensable glass vials containing the sources, and its sealing during the 2008-10 period, were made possible thanks to the expert hand of VT's Chemistry glassblower T. Wertalik. The development of the improved radon-loading system, and the germanium counter instrumentation, are hosted by the INFN Napoli laboratories in University Federico II. We also thank P. Lombardi, R. Pompilio and A. Formozov for their help and availability with the PMT characterization setups.

\section*{References}
\footnotesize
\bibliography{Radon}
\normalsize
\end{document}